\documentstyle[prl,aps,twoside,floats,epsf]{revtex}
\begin{document}
\draft
\twocolumn[\hsize\textwidth\columnwidth\hsize\csname
@twocolumnfalse\endcsname
\draft 
\title{
Semi-classical theory of magnetic quantum oscillations in 
a two-dimensional multiband  canonical Fermi liquid. 
}

\author{ A.S. Alexandrov$^1$ and A.M. Bratkovsky$^2$ }
\address{
$^{(1)}$
Department of Physics, Loughborough University, LE11 3TU, United Kingdom\\
$^{(2)}$%
Hewlett-Packard Laboratories, 1501 Page Mill Road, Palo Alto, California
94304}
\date{ July 20, 2000 }
\maketitle
\begin{abstract}
The semi-classical Lifshitz-Kosevich (LK) description of quantum
oscillations is extended to a multiband two-dimensional Fermi liquid
with a constant  number of electrons. 
The amplitudes of novel  oscillations with combination frequencies,
recently   predicted  and observed experimentally,
are analytically derived and compared  with the single-band amplitudes. 
The combination amplitudes decay with temperature exponentially
faster than the standard harmonics, and this provides a valuable tool
for their experimental identification.

\end{abstract}
\pacs{71.25.Hc, 71.18.+y, 71.10.Ay, 71.70.Di, 71.10.Pm}
\vskip 2pc ]%} 
% end \twocolumn[...]
\narrowtext

\bigskip 

It has been shown \cite{alebra} that the magnetic quantum oscillations
in a multiband $2D$  metal with a fixed electron density (canonical
ensemble ($CE$)) are qualitatively different from those in  an open
system where the chemical potential is fixed (grand canonical ensemble ($GCE$%
)). There is a mechanism for different bands to 'talk' to each other in $CE$
producing a dHvA signal with the combination  frequencies, $\ f=f_{1}+f_{2}$
\cite{alebra} and $\ f=f_{1}-f_{2}$\cite{alebra1,nakano} in addition to  the
ordinary dHvA frequencies, $f_{1,2}$, of the individual bands in $GCE.$
Numerical studies of novel oscillations showed that their amplitudes are
comparable with the standard components, and 
they are  robust with respect to a  {\em background} (non-quantized)
density of states \cite{alebra1}.

The novel frequencies have been recently observed \cite{she} in 
quantum well structures. These additional components in the dHvA frequency
spectrum of low-dimensional metals may provide a unique information on the
Fermi surface and carrier density if detailed analytical theory is
available. In this paper 
we develop such a theory in the framework of  the semiclassical LK approach 
\cite{lk}. The equations (\ref{eq:Ft}) and (\ref{eq:Ftver}) are the
main results of the present paper.

We first derive a convenient expression for a multiband
two-dimensional thermodynamic potential in an external magnetic field $H$,
\begin{equation}
\Omega=-T \int d\epsilon N(\epsilon,B)\ln\left
[1+\exp\left({\mu-\epsilon\over{T}}\right)\right]
\end{equation}
where 
\begin{equation}
N(\epsilon,B)=\sum_{\alpha}\sum_{n=0}^{\infty}\rho_{\alpha}\omega_{\alpha}
\delta(\epsilon-\epsilon_{\alpha n})
\end{equation}
is the quantized density of states,
 $\rho_\alpha$ is the zero-field density of states in the band $\alpha$,
$\epsilon_{\alpha n}=\Delta_{\alpha
 0}+\omega_{\alpha}(n+1/2)+g_{\alpha}\sigma \mu_{B}B$, 
$\omega_{\alpha}= eB/m_{\alpha}$ the cyclotron frequency with the cyclotron mass 
$m_{\alpha}$, $B=H+4\pi M$ the magnetic field, $\Delta_{\alpha 0} $
 the band edge in zero magnetic field, 
$\mu$
the chemical potential, 
$g_{\alpha}$ the electron $g$-factor, $\sigma=\pm 1/2$, $\mu_{B}$ the Bohr magneton, 
and $\hbar=c=k_{B}=1$. The band index $\alpha$ includes the electron spin. 
In actual experiments on 2DEG $B\approx H$ and magnetic coupling
between subbands was negligible\cite{she}.

By applying 
the Poisson formula \cite{sho} to the sum over $n$ in the
thermodynamic potential
\begin{equation}
\Omega=-T\sum_{\alpha}\rho_{\alpha}\omega_{\alpha}\sum_{n=0}^{\infty}
\ln\left[1+\exp\left({\mu_{\alpha}-\omega_{\alpha}(n+1/2)\over{T}}\right)\right]
\end{equation}
with $\mu_{\alpha}=\mu-\Delta_{\alpha}$ and
 $\Delta_{\alpha}=\Delta_{\alpha 0}+g_{\alpha}\sigma\mu_{B}B$,
we obtain
\begin{equation}
\Omega=\Omega_{0}+\tilde{\Omega},
\end{equation}
where
\begin{equation}
\Omega_{0}=-T\int_{0}^{\infty} d\epsilon \sum_{\alpha}\rho_{\alpha} 
\ln \left[1+\exp\left({\mu_{\alpha}-\epsilon\over{T}}\right)\right]
\end{equation}
is the 'classical' part. In $GCE$ it  does not oscillate as a function
of $1/B$, and contains the contribution due to spin susceptibility
(Pauli paramagnetism).
 The  second  part is
\begin{equation}
\tilde {\Omega}={1\over{24}}\sum_{\alpha}\rho_{\alpha}\omega_{\alpha}^2 
+2\sum_{\alpha}\sum_{r=1}^{\infty} A_{\alpha}^{r} 
\cos\left({rf_{\alpha}\over{B}} +\phi_{\alpha}^{r}\right),
\end{equation}
where the first  term produces the Landau diamagnetism and the 
second oscillatory term is responsible for the de Haas-van Alphen
effect. It is small  compared with the 'classical' part 
as $\tilde{\Omega}/\Omega_{0} \sim (\omega/\mu)^{2}$. The Fourier components
appear 
with frequencies $rf_{\alpha}\equiv rS_{\alpha}/e$, where $S_{\alpha}=
2\pi m_{\alpha}(\mu-\Delta_{\alpha 0})$ 
is the Fermi surface zero-field cross-section. The amplitudes of the
Fourier harmonics are 
\begin{equation}
A_{\alpha}^{r}= {T\rho_{\alpha}\omega_{\alpha}
\over{2r\sinh(2\pi^2rT/\omega_{\alpha})}},
\end{equation}
and the phase $\phi_{\alpha}^{r}=r \pi (1+g_{\alpha}\sigma)$. 

Differently from $GCE$, the chemical potential oscillates in
$CE$. Hence, the 'classical'  
part of $\Omega$ contributes to  oscillations as well. The relevant  
thermodynamic potential of $CE$ is the free energy $F=\Omega+\mu N$,
with a fixed  number 
of electrons,
$N=-\partial \Omega/\partial \mu$. At low temperatures we find 
\begin{equation}
\Omega_{0}=
-\sum_{\alpha}\rho_{\alpha}\mu_{\alpha}^2/2, 
\end{equation}
so that
\begin{equation} 
\mu={1\over \rho}\left(N+\sum_{\alpha}
\rho_{\alpha}\Delta_{\alpha}+{\partial \tilde{\Omega}\over{\partial
\mu}}\right), 
\label{eq:mu}
\end{equation}
where $\rho=\sum_{\alpha}\rho_{\alpha}$ is the total density of states.
Substituting this expression into $\Omega_{0}$, Eq.(8), we  obtain
\begin{equation}
F=F_{0}+\tilde{F},
\end{equation}
where the smooth non-oscillatory part of the free energy is given by
\begin{equation}
F_{0}={1\over{2 \rho}}\left( N+\sum_{\alpha}\rho_{\alpha} \Delta_{\alpha}\right)^2
-{1\over{2}}\sum_{\alpha}\rho_{\alpha}\Delta_{\alpha}^2,
\end{equation}
while the most essential oscillatory part is
\begin{equation}
\tilde{F}=
 \tilde{\Omega}-{1\over{2\rho}}\left({\partial 
\tilde{\Omega}\over{\partial \mu}}\right)^2.
\label{eq:Ft}
\end{equation}
In a more  explicit form we obtain
\begin{eqnarray}
\tilde{F}&=&{1\over{24}}\sum_{\alpha}\rho_{\alpha}\omega_{\alpha}^2+
2\sum_{\alpha,r} A_{\alpha}^{r} 
\cos\left({rf_{\alpha}\over{B}} 
+\phi_{\alpha}^{r}\right)\cr
&-& 4 \sum_{\alpha,\alpha',r,r'}
C^{r r'}_{\alpha \alpha'}
\sin\left({rf_{\alpha}\over{B}} +\phi_{\alpha}^{r}\right)
 \sin\left({r'f_{\alpha'}\over{B}} +\phi_{\alpha'}^{r'}\right).
\label{eq:Ftver}
\end{eqnarray}
It is the last term, which yields combination  Fourier harmonics with the 
combination
frequencies $f=
rf_{\alpha}\pm r'f_{\alpha'}$. Their amplitudes, 
\begin{equation}
C^{r r'}_{\alpha \alpha'}=2 \pi^2 {rr'A_{\alpha}^{r} A_{\alpha'}^{r'}
\over{\rho \omega_{\alpha}\omega_{\alpha'}}} 
\end{equation}
are comparable with the standard single-band harmonics at low temperatures, 
$ T<\omega_{\alpha}/2 \pi^2r$,
as also found in the numerical analysis \cite{alebra,alebra1} and in the experiment \cite{she}.
For example, the ratio of the combination amplitude to a single-band one    
for  $r=r'=1$  and $T=0$ in a metal with two parabolic bands ($\rho_{\alpha}=m_{\alpha}/2\pi$) is
\begin{equation}
{2C^{1 1}_{\alpha \alpha'}\over{A_{\alpha}^{1}}}={m_{\alpha}\over{m_{\alpha}+m_{\alpha'}}}.
\end{equation}
Differentiating $\tilde{\Omega}$  in $GCE$ and $\tilde{F}$ in $CE$ with respect 
to $H$ one obtains the ratio of the
combination and single band amplitudes of magnetization as

\begin{equation}
{M^{1 1}_{\alpha \alpha'}\over{M^{1}_{\alpha}}}={2C^{1 1}_{\alpha \alpha'}
\over{A_{\alpha}^{1}}} {f\over f_\alpha}={m_{\alpha}\over{m_{\alpha}
+m_{\alpha'}}}
{f\over{f_{\alpha}}},
\label{eq:M11}
\end{equation}
Differentiating twice the ratio in susceptibility is ($T=0$)
\begin{equation}
{\chi^{1 1}_{\alpha \alpha'}\over{\chi_{\alpha}^{1}}}=
{2C^{1 1}_{\alpha \alpha'}\over{ A^{1}_{\alpha}}}\left(f\over
f_\alpha\right)^2={m_{\alpha}\over{m_{\alpha} + m_{\alpha'}}}
\left({f\over{f_{\alpha}}}\right)^2.
\end{equation}
Last two ratios may be even larger than unity for the 'plus' combination harmonic 
($f=f_{\alpha'}+f_{\alpha}$)
while the 'minus' one ($f=f_{\alpha'}-f_{\alpha}$) is suppressed in magnetization and 
susceptibility, Fig.1.
At higher temperatures the combination harmonics are exponentially
small compared with 
the single-band ones, as shown in Fig.1. 

We note that according to Eq.(\ref{eq:Ftver})  the difference between the second
(and higher) Fourier amplitudes for open and closed systems should be
seen even in a simplest single-band metal. On the other hand in three
(and higher)- band metals a mixture  of more than two
different frequencies can be observed due to non-parabolic band
dispersion giving rise  to cubic and higher powers 
of the chemical potential in the expression for $\Omega_{0}$, Eq.(8).
In very high magnetic fields the usual magnetic breakdown and the
non-linear field dependence of magnetic subbands due to
non-parabolicity of the band dispersion could also lead to combination
frequencies\cite{han}.

In conclusion we have developed the analytical semiclassical theory
of magnetic quantum oscillations in  multi-band two-dimensional metals.
We have found the amplitudes of the novel combination Fourier harmonics, which are comparable
with the single-band harmonics at low temperatures and exponentially small
 at higher temperatures.
Their frequencies and the temperature dependence of  amplitudes
provide additional information 
on the band structure and carrier densities of a multiband canonical Fermi liquid.
Essentially different temperature dependence of the
combination amplitudes compared with the standard 
harmonics, Fig.1, should allow to distinguish them  experimentally.

We thank A.P. Levanyuk for enlightening comments on the
relationship between different statistical ensembles.

%%%%%%%%%%%%%%%%%%%%%%%%%%%%%%%%%%%%%%%%%%%%%%%%%%%%%%%%%%%%%%%%%%%%%%%%%
%  FIG. 1
%%%%%%%%%%%%%%%%%%%%%%%%%%%%%%%%%%%%%%%%%%%%%%%%%%%%%%%%%%%%%%%%%%%%%%%%%
\begin{figure}[t]
\epsfxsize=3.4in
\epsffile{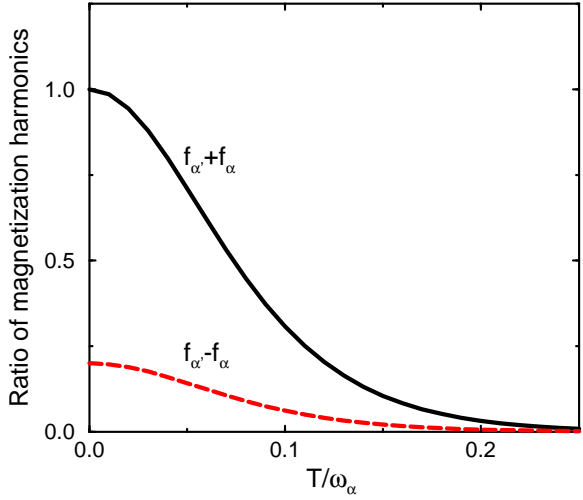}
\vspace{.02in}
\caption{The relative values of $f_{\alpha'}+f_\alpha$ and
$f_{\alpha'}-f_\alpha$ combination~Fourier components of magnetization for 
$m_{\alpha'}=1.5m_{\alpha}$~and~$f_{\alpha'}=1.5 f_{\alpha}$.
The ratio  for the parabolic bands is 
$ M^{1 1}_{\alpha \alpha'}/{M^{1}_{\alpha}}$ $= 2\pi^2Tm_{\alpha'}(f_{\alpha'} \pm
f_\alpha)/$ $[f_\alpha(m_\alpha+m_{\alpha'})\omega_\alpha
\sinh(2\pi^2T/\omega_{\alpha'})]$.}
\end{figure}

\end{document}